\documentclass{article}
\usepackage{epsfig}
\usepackage{amsmath}
\usepackage{amsfonts}
\usepackage{amssymb}
\usepackage{euscript}
\usepackage{graphicx}
\usepackage[cp1251]{inputenc}
\usepackage[russian]{babel}

\begin{document}
\sloppy

\begin{center}
{{\Huge\bfseries  The critical temperature of superconductor and its electronic specific heat \\}}
\end{center}

\begin{center}
{\Large

\itshape{ B.V.Vasiliev}}
\end{center}

\section{Introduction}

The task of the critical parameters of a superconductor calculating is reducing to the determining of their relationship to other properties of its electronic system.

 It was shown \cite{BV1}, that the assemble of superconducting carriers can be considered as a Bose-Einstein condensate. At $T=0$ the density of particles in this superconducting condensate
\begin{equation}
n_0\approx\left(\frac{m_e}{\pi^2\alpha \hbar^2}\Delta_0\right)^{3/2}.\label{n0}
\end{equation}
Where $\Delta_0$ is the size of energetic gap in a superconductor at $T=0$,
$\alpha=1/137$ is the fine structure constant.

At $T_c$ all superconducting carriers jump on the levels lying above the Fermi level. Since this process of "evaporation" $~$ depends on the specific heat of the electron gas, it allows to associate the critical temperature of the superconductor with its electronic specific heat.

\section{The electron states density and specific heat}
Let us consider the process of heating the electron gas in a superconductor.
At a  heating, the electrons from levels slightly below the Fermi-energy are raised to higher levels. As a result, the levels closest to the Fermi level, from which at low temperature electrons was forming bosons, become vacant.

At critical temperature $T_c$, all electrons from the levels of energy bands from $\mathcal{E}_F-\Delta$ to  $\mathcal{E}_F$ are moved to higher levels (and the gap collapses).At this temperature superconductivity is destroyed completely.

This band of energy can be filled by $N_\Delta$ particles:
\begin{equation}
N_\Delta=2\int_{\mathcal{E}_F-\Delta}^{\mathcal{E}_F}
F(\mathcal{E})D(\mathcal{E})d\mathcal{E}\label{ne}.
\end{equation}
Where  $F(\mathcal{E})=\frac{1}{e^{\frac{\mathcal{E}-\mu}{\tau}}+1}$ is the Fermi-Dirac function, $D(\mathcal{E})$ is  number of states per an unit energy interval, deuce front of the integral arises from the fact that there are two electron at each energy level.

To find the density of states $D(\mathcal{E})$,  one needs to find the difference in energy of the system at $T=0$ and finite temperature:
\begin{equation}
\Delta \mathcal{E} =\int_0^\infty F(\mathcal{E})\mathcal{E}D(\mathcal{E})d\mathcal{E}-\int_0^{\mathcal{E}_F} \mathcal{E}D(\mathcal{E})d\mathcal{E}\label{dE}.
\end{equation}
After  calculations \cite{Kit}, the electron level density can be expressed  through  the Fermi-energy:
\begin{equation}
D({E}_F)=\frac{dn_e}{d\mathcal{E}_F}=\frac{3n_e}{2\mathcal{E}_F}=\frac{3\gamma}{k^2\pi^2},\label{d}
\end{equation}
where
\begin{equation}
\gamma=\frac{\pi^2  k^2 n_e}{2\mathcal{E}_F}\label{ga}
\end{equation}
is the Sommerfeld constant.

To consider the interaction in the electron gas, it is commonly accepted to establish an effective electron mass $m_\star$, and to record the Fermi energy in the form
\begin{equation}
\mathcal{E}_F=(3\pi^2)^{2/3}\frac{\hbar^2}{2m_\star}n_e^{2/3}.\label{eF0}
\end{equation}
In this case the Sommerfeld constant:
\begin{equation}
\gamma=\left(\frac{\pi}{3}\right)^{3/2}\left(\frac{k}{\hbar}\right)^2 m_\star n_e^{1/3}\label{gz}
\end{equation}
and thus
\begin{equation}
\mathcal{E}_F \simeq \left(\frac{3\hbar^3\gamma}{2^{1/2}k^2 m_\star^{3/2}}\right)^2.\label{eF}
\end{equation}

At the using of similar arguments, we can find the number of electrons, which populate the levels in the range from $\mathcal{E}_F-\Delta$ to $\mathcal{E}_F$. For an unit volume of substance Eq.(\ref{ne}) can be rewritten as:
\begin{equation}
n_\Delta=2kT\cdot D(\mathcal{E}_F)\int_{-\frac{\Delta_0}{kT_c}}^0  \frac{dx}{(e^x +1)}. 
\end{equation}

At taking into account that for superconductors $\frac{\Delta_0}{kT_c}=1.76$, as a result of numerical integration we obtain
\begin{equation}
\int_{-\frac{\Delta_0}{kT_c}}^0  \frac{dx}{(e^x +1)}=\left[x-ln(e^x+1)\right]_{-1.76}^0\approx 1.22 . 
\end{equation}
Thus, the density of electrons, which the heat throws up above the Fermi level in a metal at temperature $T = T_c$ is
\begin{equation}
n_e(T_c)\approx 2.44 \left(\frac{3\gamma}{k^2\pi^2}\right)kT_c.
\end{equation}
Where the Sommerfeld constant $\gamma$ is  related to the volume unit  of the metal. \footnote{The values of the electronic specific heat of superconductors of type I recalculated to the volume unit   are shown in the middle column Tab.(\ref{2t}.4).}

\section{The Sommerfeld constant and critical temperature superconductors}
\subsection{The type-I superconductors}

The values of the Sommerfeld constant, calculated for the free electron gas Eq.(\ref{gz}), can be considered as consistent on order of magnitude with the  experimentally measured values for metals, which are type-I superconductors (see Table (\ref{cz1}.1)).

\bigskip

Table (\ref{cz1}.1).

\label{cz1}
\begin{tabular}{||c|c||}\hline\hline
  superconductor&$\gamma_{calc}/\gamma_{measured}$\\\hline
  Cd &$1.34$\\
  Zn &$1.11$\\
  Ga &$1.65$\\
  Al &$0.66$\\
  Tl &$0.87$\\
  In &$0.72$\\
  Sn &$0.77$\\
  Hg &$0.54$\\
  Pb &$0.47$\\ \hline\hline
\end{tabular}

\bigskip
Therefore, in a rough approximation, the effective mass of the electron in these metals can be considered approximately equal to the free electron mass:
\begin{equation}
\frac{m_\star}{m_e}\approx 1.
\end{equation}

This is a consequence of the fact that all metals - type-I superconductors - have completely filled inner shells
 (see. Table(\ref{I}.2)).

\bigskip
Table (\ref{I}.2).

\label{I}
\begin{tabular}{||c|c||}\hline\hline
  superconductors&electron shells\\\hline
  $Al$ &$~~~~~2d^{10}~3s^2~3p^1$\\
  $Zn$ &$3d^{10}~4s^2$\\
  $Ga$ &$~~~~~3d^{10}~4s^2~4p^1$\\
  $Cd$ &$4d^{10}~5s^2$\\
  $In$ &$~~~~~4d^{10}~5s^2~ 5p^1$\\
  $Sn$ &$~~~~~4d^{10}~5s^2~5p^2$\\
  $Hg$ &$5d^{10}~ 6s^2$\\
  $Tl$ &$~~~~~5d^{10}~6s^2~ 6p^1$\\
  $Pb$ &$~~~~~5d^{10}~6s^2 ~6p^2$\\\hline\hline
\end{tabular}

\bigskip

For this reason, the electrons from the inner shells should not affect the process of thermal excitation of free electrons and their movement.

These calculations make it possible to link directly the critical temperature superconductor with the experimentally measurable parameter of a solid-state - its electronic specific heat.

The density of superconducting carriers at $T=0$ has been calculated earlier (Eq.(\ref{n0})).

\begin{figure}
\hspace{1.5cm}
\includegraphics[scale=0.5]{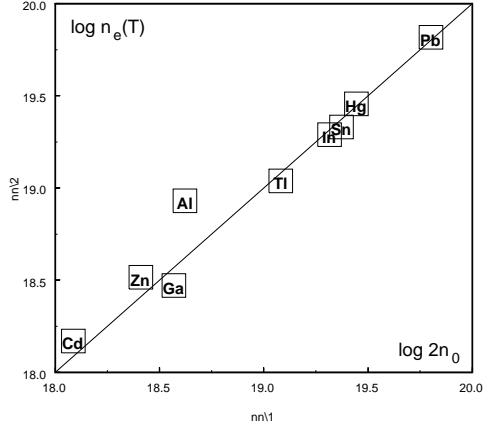}
\caption {The comparison of the number of superconducting carriers at $T=0$ with the number of thermally activated electrons at $T=T_c$.}\label{n0Ne}
\end{figure}

The comparison of the values $n_0$ and $n_e(T_c)$ is given in the Table (\ref{nna}.3) in Fig.(\ref{n0Ne}).
(The necessary data for superconductors are taken from the tables (\cite{Pool}), (\cite{Kett})).

\bigskip
Table (\ref{nna}.3).

\begin{tabular}{||c|c|c|c||}\hline\hline
  superconductor&$n_0$&$n_e({T_c})$&$2n_0/n_e(T_c)$\\\hline
  Cd &$6.11\cdot 10^{17}$&$1.48\cdot 10^{18}$&0.83\\
  Zn &$1.29\cdot 10^{18}$&$3.28\cdot 10^{18}$&0.78\\
  Ga &$1.85\cdot 10^{18}$&$2.96\cdot 10^{18}$&1.25\\
  Al &$2.09\cdot 10^{18}$&$8.53\cdot 10^{18}$&0.49\\
  Tl &$6.03\cdot 10^{18}$&$1.09\cdot 10^{19}$&1.10\\
  In &$1.03\cdot 10^{19}$&$1.94\cdot 10^{19}$&1.06\\
  Sn &$1.18\cdot 10^{19}$&$2.14\cdot 10^{19}$&1.10\\
  Hg &$1.39\cdot 10^{19}$&$2.86\cdot 10^{19}$&0.97\\
  Pb &$3.17\cdot 10^{19}$&$6.58\cdot 10^{19}$&0.96\\ \hline\hline
\end{tabular}
\label{nna}
\bigskip

From the data obtained above, one can see that the condition of destruction of superconductivity after heating  for  superconductors of type I can really be written as the equation:
\begin{equation}
{n_e}{({T_c})}=2{n_0}\label{nn}
\end{equation}

    Eq.(\ref{nn}) gives us the possibility to express the critical temperature of the superconductor through its Sommerfeld constant :
\begin{equation}
\Delta_0\simeq C\gamma^2\label{tcc},
\end{equation}
where the constant
\begin{equation}
C\simeq 35\frac{\pi^2}{k} \left[\frac{\alpha\hbar^2}{k m_e}\right]^3\approx 6.5\cdot 10^{-22}\frac{K^4 cm^6}{erg}\label{tc2}.
\end{equation}
or at Eq.(\ref{eF}),
\begin{equation}
\Delta_0 \simeq 8\pi^2\alpha^3 \mathcal{E}_F\label{DEF}.
\end{equation}

\begin{figure}
\includegraphics[scale=.5]{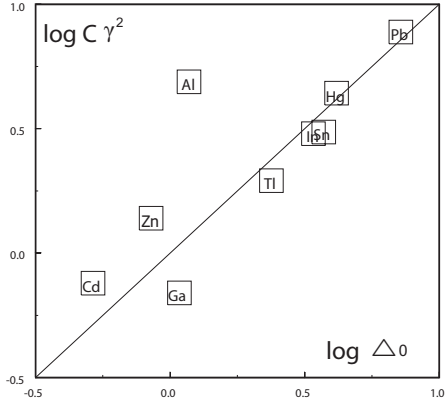}
\caption {The comparison of the calculated values of critical temperature superconductors with measurement data. On abscissa the measured values of the critical temperature superconductors of type I are shown, on the vertical axis the function $C\gamma^2$  defined by Eq.(\ref{tcc}) is shown.}\label{gamma}
\end{figure}

The comparison of the temperature calculated by Eq.({\ref{tcc}})  (corresponding to complete evaporation of electrons with energies in the range from $\mathcal{E}_F-\Delta_0$ up to $\mathcal{E}_F$) and the experimentally measured critical temperature superconductors is given in Table (\ref{2t}.4) and in Fig.(\ref{gamma}).
\bigskip

\begin{tabular}{||c|c|c|c|c||}\hline\hline
  superconductors&$T_c$(measur),K&$\gamma,\frac{erg}{cm^3K^2}$&$T_c$(calc),K&$\frac{T_c (calc)}{T_c(meas)}$\\
  &&&Eq.(\ref{tcc})&\\\hline
  Cd &$0.517$&532&$0.77$&1.49\\
  Zn &$0.85$&718&$1.41$&1.65\\
  Ga &$1.09$&508&$0.70$&0.65\\
  Tl &$2.39$&855&$1.99$&0.84\\
  In &$3.41$&1062&$3.08$&0.90\\
  Sn &$3.72$&1070&$3.12$&0.84\\
  Hg &$4.15$&1280&$4.48$&1.07\\
  Pb &$7.19$&1699&$7.88$&1.09\\ \hline\hline
\end{tabular}\label{2t}

Table (\ref{2t}.4).
\bigskip

\subsection{The estimation of properties of type-II superconductors}

The situation is different in the case of type-II superconductors.

In this case the measurements show that these metals have the electronic specific heat on an order of magnitude greater than the calculations based on free electron gas give.

 The peculiarity of these metals associated with the specific structure of their ions.
  They are transition metals with unfilled inner d-shell (see Table\ref{mII}).

It can be assumed that this increasing of the electronic specific heat of these metals should be associated with a characteristic interaction of free electrons with the electrons of the unfilled d-shell.

\bigskip

Таблица (\ref{mII})

\begin{tabular}{||c|c||}\hline\hline
  superconductors &electron shells\\\hline
  $Ti$ &$3d^2 ~4s^2$\\
  $V$ &$3d^3  ~4s^2$\\
  $Zr$ &$4d^2~ 5s^2$\\
  $Nb$ &$4d^3 ~5s^2$\\
  $Mo$ &$4d^4~ 5s^2$\\
  $Tc$ &$4d^5 ~5s^2$\\
  $Ru$ &$4d^6~ 5s^2$\\
  $La$ &$5d^1 ~6s^2$\\
  $Hf$ &$5d^2 ~6s^2$\\
  $Ta$ &$5d^3 ~6s^2$\\
  $W$ &$5d^4~6s^2$\\
  $Re$ &$5d^5 ~6s^2$\\
  $Os$ &$5d^6 ~6s^2$\\
  $Ir$ &$5d^7~6s^2$\\\hline\hline
\end{tabular}\label{mII}

\bigskip

An electron of conductivity at the moving with velocity v has the kinetic energy
\begin{equation}
\mathcal{E}_k=\frac{m_e v^2}{2}.
\end{equation}
In a contact with d-electron shell, the electron of conductivity creates on it a magnetic field
\begin{equation}
H\approx \frac{e}{r_c^2}\frac{v}{c}.
\end{equation}
The magnetic moment of d-electron is approximately equal to the Bohr magneton, so the energy of their interaction will be approximately equal to:
\begin{equation}
\mathcal{E}_\mu \approx \frac{e^2}{2r_c}\frac{v}{c}.
\end{equation}
This leads to the fact that an  electron of conductivity  with its movement would be to spend energy on the orientation of the magnetic moments of d-electrons.
It should increase its effective mass to:
\begin{equation}
m_\star \approx m_e\left(1+\frac{\mathcal{E}_\mu}{\mathcal{E}_k}\right)\approx m_e\frac{\mathcal{E}_\mu}{\mathcal{E}_k}.
\end{equation}

Thus, the electronic specific heat in the transition metals should be composed from two parts. The first term - the same as in type-I superconductors. It is the heating of collectivized electrons - the increasing of their kinetic energy. The second term is associated with the excitation of additional potential energy of the moving collectivized electron, i.e. interaction with the magnetic moments of electrons unfilled d-shell. \footnote{Although estimates show that the second mechanism is stronger, to see its work is not simple. Only at temperature below 0.5K, this mechanism leads to the square-root dependence of electronic specific heat on temperature: $c_{el} \sim\sqrt{T}$.
But simulation shows that at higher temperature the Debye cubic lattice term will mask it, and the total heat capacity dependence of transition metals are similar to the linear in T.}

    The interaction with d-electrons increases the effective mass of free electrons. Therefore, in considering this process should take into account the factor:
    \begin{equation}
\frac{\mathcal{E}_k}{\mathcal{E}_\mu+\mathcal{E}_k}\approx \frac{1}{\alpha^{3/2}}\sqrt{\frac{2\Delta_0}{m_e c^2}}.
\end{equation}
At taking into account the effective mass of electron instead Eq.(\ref{DEF})
for the type II superconductors we become
\begin{equation}
\Delta_0\simeq 8 \pi^2\alpha^3 \mathcal{E}_F \left(\frac{\mathcal{E}_k}{\mathcal{E}_\mu+\mathcal{E}_k}\right)
\end{equation}
or
\begin{equation}
\Delta_0\simeq C \gamma^2\left(\frac{\mathcal{E}_k}{\mathcal{E}_\mu+\mathcal{E}_k}\right)\simeq \frac{2}{m_e c^2}\left(\frac{6\pi\alpha^{3/4}\hbar^3}{k^2m_e^{3/2}}\right)^4\gamma^4.\label{delta2}
\end{equation}

\begin{figure}
\includegraphics[scale=.5]{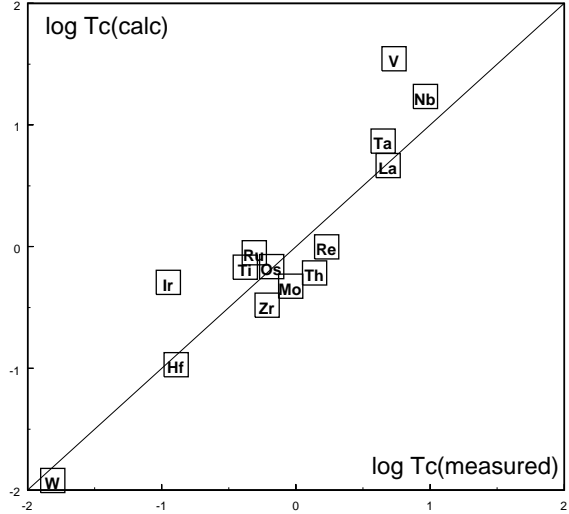}
\caption {The comparison of the calculated values of critical temperatures for type II superconductors (Eq.(\ref{delta2})) with measurement data.}
\label{Tc2g}
\end{figure}

 The comparing of the results of these calculations with the measurement data (Fig.(\ref{Tc2g})) shows that for the majority of type II superconductors the estimation Eq.(\ref{delta2}) can be considered quite satisfactory.

\bigskip

\subsection{Alloys and high-temperature superconductors}
To understand the mechanism of high temperature superconductivity is important to establish whether the high-Tc ceramics are the I or II type superconductors, or they are a special class of superconductors.

 To address this issue, you can use the established relation of critical parameters with the electronic specific heat and the fact that the specific heat of superconductors I and II types are differing considerably, because in the I type superconductors the energy of heating  spends to increase the kinetic energy of the electron gas, while in the type II superconductors  it is spending  additionally on the polarization of electrons of the unfilled d-shell.

There are some difficulty on this way - one d'not known confidently the Fermi-energy and the density of the electron gas in high-temperature superconductors.
However, the density of atoms in metal crystals does not differ too much. It
facilitates the solution of the problem of  a distinguishing of  I and II types superconductors at using of  Eq.(\ref{tcc}).

For the I type superconductors at using this equation we get the quite satisfactory estimation of the critical temperature (as was done above (see Fig.(\ref{gamma})). For the type-II superconductors this  assessment gives overestimated value due to the fact that their specific heat has additional term associated with the polarization of d-electrons.

\begin{figure}
\includegraphics[scale=.4]{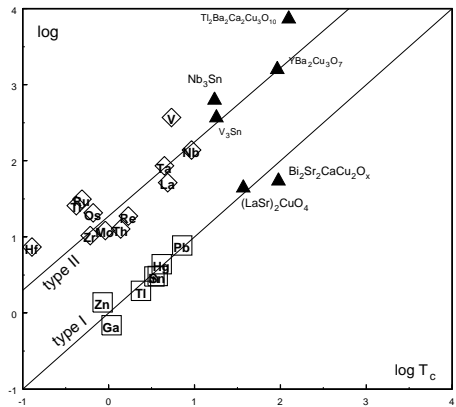}
\caption {The comparison of the calculated parameter $C\gamma^2$ with the measurement critical temperatures of elementary superconductors and some superconducting compounds.}\label{gamma2}
\end{figure}

Indeed, such analysis gives possibility to share all of superconductors into two groups, as is evident from the figure (\ref{gamma2}).

It is generally assumed to consider alloys $Nb_3Sn$ and $V_3Si$ as the type-II superconductors. The fact seems quite normal that they are placed in close surroundings of Nb.
Some excess of the calculated critical temperature over the experimentally measured value for ceramics $Ta_2Ba_2Ca_2Cu_3O_{10}$ can be attributed to the fact that the measured heat capacity may make other conductive electrons, but nonsuperconducting elements (layers) of ceramics. It is not a news that it, as well as ceramics $YBa_2Cu_3O_7$, belongs to the type II superconductors. However, ceramic $(LaSr)_2CuO_4$ and $Bi_2Sr_2CaCu_2 O_x$, according to this figure should be regarded as type-I superconductors, which is somewhat unexpected.

\section{Conclusion}

The above analysis shows that the critical temperature of the superconductor is related to the Sommerfeld constant and for type-I superconductors the critical temperature is proportional to the first degree of its Fermi energy. Estimates for type II superconductors lead to a quadratic dependence of the critical temperature of the Fermi energy, but the relationship of the critical temperature and critical field is the same for different types of superconductors \cite{BV0}. While the above analysis does not detect any effect of electron-phonon interaction on the phenomenon of superconductivity.
That and the fact that the superconductivity in metals should be considered as a phenomenon of Bose-Einstein condensation in a system of electron gas \cite{BV1} indicate that superconductivity is the effect of electronic interactions. In our opinion, for the phenomenon of superconductivity should be responsible ordering zero-point oscillations in the metal electron gas \cite{BV0}.

\end{document}